\documentclass[utf8]{frontiersSCNS} 
\usepackage{url,hyperref,lineno,microtype,subcaption}
\usepackage[onehalfspacing]{setspace}
\def\apss{Astroph.   Space Sci.}
\def\apj{Astroph. J.}\def\aj{Astron. J.}\def\apjl{Astroph. J. Lett.}
\def\mnras{Mon. Not. R. Astron. Soc.}
\def\aap{Astron. Astroph. }\def\apjs{Astroph. J. Suppl. Ser.}
\def\nat{Nature}\def\araa{Ann. Rev. Astron. Astroph.}
\defcitealias{marzianietal16}{M16}
\defcitealias{marzianietal16a}{M16a}

\def\keyFont{\fontsize{8}{11}\helveticabold }
\def\firstAuthorLast{Marziani {et~al.}} 
\def\Authors{P. Marziani\,$^{1,*}$, C. A. Negrete\,$^{2,*}$, D. Dultzin\,$^{2,*}$, M. L. Mart\'{i}nez-Aldama\,$^{3,*}$,  A. Del Olmo\,$^{3,*}$,  M. D'Onofrio\,$^{4,*}$, G. M. Stirpe\,$^{5,*}$}
\def\Address{$^{1}$ INAF, Osservatorio Astronomico di Padova, Italy \\
$^{2}$\ Instituto de Astronom\'{i}a, UNAM, Mexico\\
$^{3}$\ Instituto de Astrof\'{i}sica de Andaluc\'{i}a (IAA-CSIC), Granada, Spain\\ 
$^{4}$ Universit\`a di Padova, Padova, Italia\\  
$^{5}$INAF Osservatorio Astronomico di Bologna, Italia
  }
\def\corrAuthor{Paola Marziani}
\def\corrEmail{paola.marziani@oapd.inaf.it}
\begin{document}
\onecolumn
\firstpage{1}
\title[Quasar massive ionized outflows]{Quasar massive ionized outflows traced by CIV $\lambda$1549 and [OIII]$\lambda\lambda$4959,5007} 
\author[\firstAuthorLast ]{\Authors} 
\address{} 
\correspondance{} 
\extraAuth{}

\maketitle
\begin{abstract}
\section{}
The most luminous quasars (with bolometric luminosities are $\gtrsim 10^{47}$ erg/s) show  a high prevalence of CIV $\lambda$1549 and [OIII]$\lambda\lambda$4959,5007 emission line profiles with strong  blueshifts.  Blueshifts are interpreted as due to Doppler effect and selective obscuration, and  indicate outflows occurring over a wide range of spatial scales. We found evidence in favor of the nuclear origin of the outflows diagnosed by [OIII]$\lambda\lambda$ 4959,5007. The  ionized gas mass, kinetic power, and mechanical thrust  are extremely high, and suggest widespread feedback { effects} on the host galaxies of very luminous quasars, at cosmic epochs between 2 and 6 Gyr from the Big Bang. In this mini-review we summarize   results { obtained by our group and reported in several major papers in the last few years } with an eye on challenging aspects of quantifying feedback effects in large samples of quasars. 
\tiny
 \keyFont{ \section{Keywords:}  galaxy evolution -- quasars -- feedback -- outflows -- quasars: emission lines  --quasars: supermassive black holes } 
\end{abstract}

\section{Introduction}

The broad and narrow high-ionization emission lines (HILs) in the optical and UV spectra of quasars frequently show significant blueshifts with respect to the quasar rest frame \citep[e.g.,][for some  early papers]{gaskell82,tytlerfan92,marzianietal96,corbinboroson96,zamanovetal02}. The interpretation  involves the Doppler shift of line radiation due to the emitting gas motion toward the observer, with the  part of line emitted by  receding gas suppressed by obscuration. In the following we will adhere with this interpretation (for a dissenting view see however \citealt{gaskellgoosmann13} { who posit that we see light originally emitted by gas falling toward  the black hole and backscattered toward us}), and consider [OIII]$\lambda$4959,5007 as representative of narrow high-ionization lines (HILs), and CIV$\lambda$1549 as a prototypical broad HIL. 

\section{The  quasar main sequence: contextualizing outflows at low-to-moderate $L$}


{  The diversity of quasar spectroscopic properties as found in single epoch observations of large samples  has  been organized along a quasar main sequence \citep[][]{sulenticetal00a,sulenticetal00b,marzianietal01,shenho14}.  \citet{borosongreen92} identified a first eigenvector in their sample of $\approx$ 80 Palomar-Green quasars which is associated with an anticorrelation between FWHM (H$\beta$) and a parameter measuring the prominence of FeII emission (the intensity ratio between the FeII blend at $\lambda4570$ and  H$\beta$). Along the main sequence defined by this anti-correlation, \citep{sulenticetal00a} suggested a change in properties in correspondence of FWHM H$\beta \approx$ 4000 km/s, and distinguished two populations: Population A (FWHM $\lesssim$ 4000 km/s) and B   \citep[where the B stands for broader than 4000 km/s; e.g., ][for reviews; Table 1  of \citealt{fraix-burnetetal17} summarized parameter systematic differences between the two populations]{marzianietal14,fraix-burnetetal17}}. Population A and B   have been associated with high and low accretion, respectively. 

The CIV$\lambda$1549 large blueshifts (above 1000 km/s) are a Population A phenomenon, likely associated with   a disk wind  (see Figure 7 of \citealt{sulenticmarziani15}). Population A  sources, { at low $z$ ($\lesssim 1$)},  encompass relatively low black hole mass quasars { ($\sim 10^{7} - 10^{8} \mathrm{M}_{\odot}$)}    radiating at a relatively high Eddington ratio ($\gtrsim 0.1 - 0.2$). In many ways Pop. A sources can be considered as an extension of Narrow Line Seyfert 1 (NLSy1) with moderate or strong FeII emission: the limit FWHM $\approx$ 4000 km/s (valid at  bolometric luminosity  $\log L \lesssim 47$\ [erg s$^{-1}$]) allows one to include sources with the same Balmer line profiles and intensity ratios as observed in NLSy1s.   This is not to imply that Pop. B sources (with FWHM H$ \beta \gtrsim$ 4000 km/s)  do not show evidence of outflows. Evidence of outflow is, for example, overwhelming in the prototypical Pop. B source NGC 5548 \citep{kaastraetal14}. The latest developments suggest that outflows are ubiquitous, also in forms that may not provide striking optical/UV spectral phenomenologies \citep[e.g., ][]{tombesietal15,harrisonetal14}.  Only,  outflows are more difficult to trace in Population B single-epoch spectra, as the CIV integrated line profiles are relatively symmetric. In both Pop. A and B, the CIV$\lambda$1549 line profile can be represented as a scaled H$\beta$\ profile plus an excess of blueshifted emission: the two components are assumed to be representative of a ``virialized'' low-ionization broad line region (producing a fairly symmetric and unshifted line) plus an  outflow/wind component with different physical conditions.  In Pop. B, CIV$\lambda$1549\   shows only a small blueshifted excess if  compared to H$\beta$. 

Similarly, large blueshifts of [OIII]$\lambda\lambda$4959,5007 above 250 km/s are rare in    $z$\ samples (they are real statistical outliers, called ``blue outliers'' [BOs] by \citealt[][]{zamanovetal02}) and have been preferentially found among Population A sources \citep[e.g.,][]{zamanovetal02,xuetal12,zhangetal13,craccoetal16}.  \citet{sulenticmarziani15} show the distribution of [OIII]$\lambda\lambda$4959,5007 peak shifts for the spectral types defined along the Eigenvector 1 sequence:  the prevalence of  [OIII]$\lambda\lambda$4959,5007 large blueshifts increases in Pop. A and reaches a maximum in extreme sources with FeII$\lambda4570$/H$\beta  \gtrsim 1$\ (Figure 5 of \citealt{sulenticmarziani15}; Negrete et al. 2017, in preparation).  Blueshifts of [OIII]$\lambda\lambda$4959,5007 trace larger scale outflows than CIV$\lambda$1549, outside of the broad line region (BLR). 

\section{The scenario at high $L$, and intermediate-$z$}

\begin{figure}[h!]
\begin{center}
\includegraphics[width=18.cm]{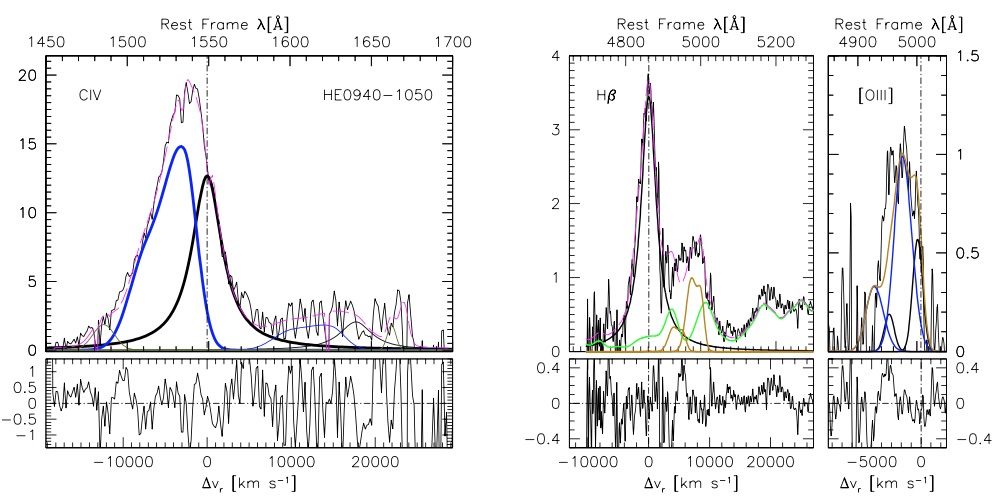}
\end{center}
\vspace{0cm}
\caption{Left panel:   CIV$\lambda$1549 emission after continuum subtraction. A scaled H$\beta$\  profile is superimposed (adapted from \citealt{sulenticetal17}). Right panel:  H$\beta$\ spectral range. The green line shows the adopted FeII template.  The full profiles of [OIII]$\lambda$$\lambda$4959,5007 are shown in orange. The thick black lines identify a symmetric unshifted emission that is dominating H$\beta$ and still contributing to CIV$\lambda$1549. The  thick blue lines trace the excess emission once the symmetric component is subtracted. The thin lines in the CIV panel show the interpretation of the HeII $\lambda$ 1640 profile assuming a symmetric  and a blue shifted component, as for CIV. The rightmost panel provides an  enlargement on the [OIII]$\lambda\lambda$4959,5007, semi-broad, boxy and blue-shifted profile (not a unique case at high $L$: e.g., \citealt{canodiazetal12}). The [OIII] profile is modeled as the sum of a moderately  blueshifted  core component (black line) and of a semi-broad component, blue-shifted by $\approx -2500$ km/s (blue line). The scale is $10^{-15}$ erg  s$^{-1}$ cm$^{-2}$ \AA$^{-1}$\ in the rest frame, as applied by \citet{sulenticetal17}.}\label{fig:1} 
\end{figure}

A recent result is that the   prevalence of large blueshifts in both  CIV$\lambda$1549 and [OIII]$\lambda$4959,5007 quasars is much higher at  high  $L$  in intermediate-$z$\ samples \citep[1 $\lesssim z \lesssim 2.5$,][M16]{marzianietal16a,coatmanetal16,zakamskaetal16,vietri17,bischettietal17}.   Blueshifts of CIV$\lambda$1549 reach several thousands km/s in Pop. A. BOs  become  much more frequent in the high $z$\ and $L$\  samples. The  [OIII] shift and FWHM distributions at high-$L$\ are remarkably different from those of low-$z$, low $L$\ samples.  Figure \ref{fig:1} shows an example of a luminous Pop. A source in the sample of \citet{sulenticetal17}: the left panel shows, overlaid to the continuum-subtracted spectrum, a decomposition of the  CIV$\lambda$1549 profile into an unshifted and symmetric  component (thick black line) and a blue shifted component (blue line). Without involving any profile decomposition (the caveats of the technique are discussed in \citealt{negreteetal14a}), it is easy to see  that about 80\%\ of the line  flux is emitted short-wards of the rest wavelength. At the same time, the H$\beta$ remains symmetric.  Fig. \ref{fig:1} (rightmost panel)  shows an enlargement of  the  [OIII]$\lambda$4959,5007 profile: the FWHM $\approx$ 3600\ km/s is extremely broad by [OIII]$\lambda$5007 standards (at low-$z$, [OIII]$\lambda$5007 FWHM is $\lesssim 1000$ km/s). The profile appears boxy, and fully displaced to the blue. 

Figure \ref{fig:1} represents the diagnostic provided by single epoch observations for quasars at intermediate-to-high $z$ (1 -- 5), with CIV$\lambda$1549 covered by optical spectrometers and [OIII]$\lambda$4959,5007 requiring near-IR spectroscopic observations. The latter   have become possible for relatively faint quasars only in recent times, with the advent of second generation IR spectrometers mounted at the foci of large aperture telescopes; two major examples are XSHOOTER at VLT and LUCI at LBT. Near IR observations provide a reliable estimate of the quasar systemic redshift if the narrow component of H$\beta$ or the [OII]$\lambda$3727 can be effectively measured. If these lines are detected above noise, then a good precision in the rest frame can be achieved, and the uncertainty is typically $\delta z \lesssim 3 \cdot 10^{-4}$. An accurate knowledge of the rest frame is not an end in itself, since an important physical parameter  such as the outflow kinetic power depends on the third power of the outflow velocity $v_\mathrm{o}$. The availability of such observations should increase dramatically in the next few years, providing useful data (even with lack of real spatial resolution)  for a better understanding of the outflow prevalence and power as a function of luminosity and cosmic epoch. { At low-$z$,  partial spatial  resolution of the [OIII]$\lambda$4959,5007 emitting regions is currently obtained (and will be more frequently obtained in the coming years)  with the use of integral-field unit spectrographs with adaptive optics \citep[e.g.,][]{crescietal15}. Observations of CIV$\lambda$1549 are and will remain challenging for sources at $z \lesssim 1.2 $\ i.e.,  for all the low-$z$ quasar population. }

\subsection{The nuclear nature of the outflow}

CIV$\lambda$1549 emission is expected to originate within a few hundred gravitational radii from the central black hole even in luminous quasars \citep{kaspietal07}. Reverberation mapping studies indicate that the CIV emitting region scaling law with luminosity is a  power-law with exponent 0.5 -- 0.7  (results by Shai Kaspi in this research topic). The nuclear nature of the CIV$\lambda$1549 outflow is not in question. 

 
The [OIII]$\lambda\lambda$4959,5007 prominence is affected by the  ``Balwdin effect'' \citep{zhangetal11},  which is perhaps the  main luminosity effect affecting all HILs \citep{dietrichetal02}.   The observations of samples covering a wide range in luminosity ($43 \lesssim \log L \lesssim 48.5$ [erg/s])  confirm the [OIII] Baldwin effect: $W \propto\  L_{5100}^{-0.26 \pm 0.03}$\ \citepalias{marzianietal16}, steeper than the ``classical'' CIV Baldwin effect. 
However,  a most intriguing result is that  sources with large  [OIII]  blueshifts (the ``blue outliers'' of \citealt{zamanovetal02}) do not follow any Baldwin effect: $W \propto\  L_{5100}^{0.050 \pm 0.066},$ as shown in Figure 4 of \citetalias{marzianietal16}. In other words, for  sources with large blueshifts, the line luminosity is proportional to the continuum luminosity. The simplest explanation is  that    [OIII]$\lambda\lambda$4959,5007 emission is due to photoionization by the nuclear continuum. 

\begin{figure}[htp!]
\begin{center}
\includegraphics[width=15cm]{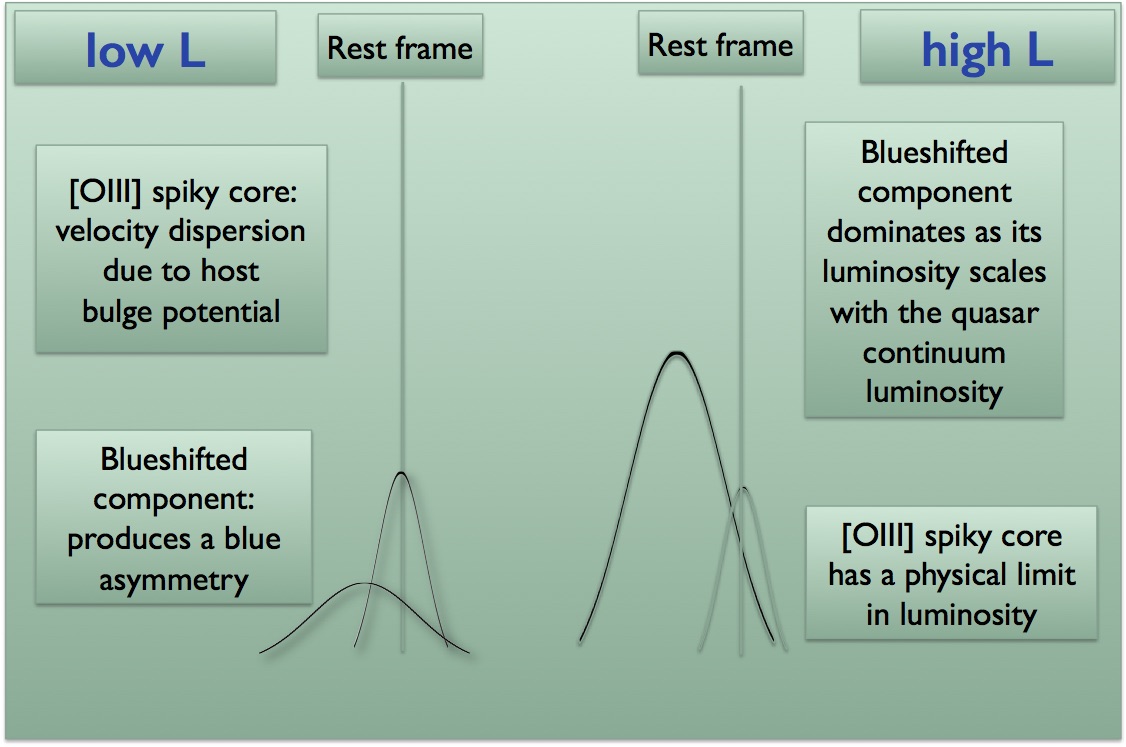} 
\end{center}
\caption{The high prevalence of large  [OIII]$\lambda\lambda$4959,5007 blueshift  at high $L$\ explained as a luminosity effect. The luminosity of the blue shifted component grows with the nuclear continuum luminosity, dominating the total [OIII] emission in very luminous quasars.  Fig. 6 of \citetalias{marzianietal16} shows the same same effect in terms of EW: the EW of the core component is lower in very luminous sources because of the strong nuclear continuum, while the EW of the blue shifted component remains constant.
\label{fig:2}}
\end{figure}

Fig. \ref{fig:2} shows a sketch  explaining   this result. At low-$L$, emission of [OIII]$\lambda$4959,5007 shows a spiky core and a prominent blueward asymmetry, especially in Pop. B sources. If the line profile is interpreted as made of a core component and a blue shifted semi broad component, then the latter component is not dominating at low-$z$ and low-$L$ unless we are considering a system radiating at high- Eddington ratio: these sources show {\em only} the semi broad component. At high-$L$ the semi-broad component becomes so luminous to overwhelm the core component whose  luminosity is expected to be upper-bound by the physical size and gas content of the host galaxy \citep{netzeretal04}. 

\begin{table*}\scriptsize
\setlength{\tabcolsep}{1pt}
\centering
\caption{Outflow physical parameters derived for CIV and [OIII]: mass of ionized gas, mass outflow rate, thrust and kinetic power   \label{tab:php}}
\begin{tabular}{|c|c|c|c|}\hline\hline
Parameter & Units & CIV & [OIII]   \\ \hline
$M^\mathrm{ion}$ & $M_{\odot}$ & $
  1.9 \cdot 10^3~L_{45} \left(\frac{Z}{5Z_{\odot}}\right)^{-1} n_{9}^{-1}  $ &
  $1 \cdot 10^7~L_{44} \left(\frac{Z}{5Z_{\odot}}\right)^{-1} n_{3}^{-1} $ \\
$ \dot{M}^\mathrm{ion}$ & $M_{\odot}$ yr$^{-1}$ &   $30  L_{45} v_\mathrm{o,5000} r^{-1}_{\rm 1pc} \left(\frac{Z}{5Z_{\odot}}\right)^{-1} n_{9}^{-1}   $ & $ 30  L_{44} v_\mathrm{o,1000} r^{-1}_{\rm 1 kpc}   \left(\frac{Z}{5Z_{\odot}}\right)^{-1} n_{3}^{-1}  $\\
$ \dot{M}^\mathrm{ion} k v_\mathrm{o} $ & g cm s$^{-2}$ &   $ 1 \cdot 10^{36}  L_{45} k v^{2}_\mathrm{o,5000} r^{-1}_{\rm 1pc}    \left(\frac{Z}{5Z_{\odot}}\right)^{-1} n_{9}^{-1}  $ & $1.9 \cdot 10^{35} L_{44} k v^{2}_\mathrm{o,1000} r^{-1}_{\rm 1 kpc}    \left(\frac{Z}{5Z_{\odot}}\right)^{-1} n_{3}^{-1} $\\

 $ \dot{\epsilon}$ & erg s$^{-1}$  &  $ 2.4 \cdot 10^{44} L_{45} k^{2} v^{3}_\mathrm{o,5000}   r^{-1}_{\rm 1pc}   \left(\frac{Z}{5Z_{\odot}}\right)^{-1} n_{9}^{-1}    $ & $ 9.6 \cdot 10^{42} L_{44} k^{2} v^{3}_\mathrm{o,1000}   r^{-1}_{\rm 1kpc}  \left(\frac{Z}{5Z_{\odot}}\right)^{-1} n_{3}^{-1}   $ \\
 \hline
\end{tabular}
\end{table*}


\section{Estimates of outflow dynamical parameters and considerations on their reliability}

Computing the kinetic power and the thrust  from single-epoch spectra is possible under several caveats and assumptions. Here we briefly recall a simplified way to estimate the mass of ionized gas, the mass outflow rate, the thrust, and the kinetic power of the outflow for collisional excited lines in photoionized gases \citep{canodiazetal12}. The formation of these authors allows to write the outflow parameters in a form independent from the covering and filling factor, provided that all emitting gas has the same density. The simplified approach is well-suited to elucidate the role of the most relevant outflow parameters in the computation of the thrust and kinetic power.  We will then consider the specific assumptions needed to apply the following relation to [OIII] and CIV measures from single epoch spectra. 

The  luminosity of any collisionally-excited line{\footnote{ While the ionic stages of C$^{3+}$\ and O$^{2+}$ are due to photoionization, the CIV and [OIII] lines are produced via collisional excitation which is dominant over recombination, as shown in detail for [OIII] in \citet{pradhannahar15}, \S 12.4.} } is   given by $ L({\rm line}) = \int _V j_\mathrm{line}~f_\mathrm{f}~dV \label{eq:}$, where $j_\mathrm{line}$ is the line emissivity per unit volume, and can be written as:  $j_\mathrm{line} = h \nu q_\mathrm{lu} n_\mathrm{e} n_\mathrm{l}$, where   $n_\mathrm{e}$ the electron density, and $n_\mathrm{l}$\ the number density of   ions at the lower level of the transition. The collisional excitation rate at electron temperature $T$\ is  $q_\mathrm{lu} = \frac{\beta}{\sqrt{T}} \frac{\Upsilon_\mathrm{lu}}{g_{l}} \exp{\left(-\frac{\epsilon_\mathrm{lu}}{kT}\right)}$, where  $g_\mathrm{l}$  is the statistical weight of the lower level, and $\Upsilon_\mathrm{lu}$\ is the effective collision strength.  The  line luminosity can be connected to the mass of ionized gas  since $M^\mathrm{ion}_\mathrm{out} \propto L_\mathrm{line}) \left(\frac{Z}{Z_{\odot}}\right)^{-1} n^{-1}  \label{eq_a5}$. Up to this point the main assumptions are: (1) constant density; (2) all emitting gas being in the ionization stage that is producing the line; (3) well defined chemical abundances.  
	 
The {mass outflow rate} at a distance $r$\  can be written as, if the flow is confined to a solid angle of $\Omega$ of volume $\frac{4}{3}\pi r^{3} \frac{\Omega}{4 \pi} $: $ \dot{M}^\mathrm{ion}_\mathrm{o} =  \rho\ \Omega r^{2} v_\mathrm{o}  = \frac{{M}^\mathrm{ion}_\mathrm{o}}{V} 	\Omega r^{2} v_\mathrm{o}    \propto  L  v_\mathrm{o}  r^{-1} $. This requires the knowledge of (4) a typical emitting region radius, and (5) the outflow velocity $v_\mathrm{o}$. Assuming a single $r$ value is already a strong simplification, especially for [OIII].    If the line emitting gas is still being accelerated (as in the CIV case), and the terminal velocity is $v_\mathrm{term} = k v_\mathrm{o}$, then the thrust should be $\propto \dot{M}k v_\mathrm{o}$ and the kinetic power of the outflow $\dot{\epsilon} \ \sim\ \frac{1}{2} \dot{M}^\mathrm{ion} k^{2} v_\mathrm{o}^{2} \propto  L_\mathrm{line}  k^{2} v_\mathrm{o}^{3}   r^{-1}$. A value of $k$=1 can be assumed for [OIII] { (as explained in  \S \ref{oiii})}.  The parameters of Table  \ref{tab:php} can be all written in the form $  \propto L_\mathrm{line} r^{-1} n^{-1} (Z/Z_{\odot})^{-1}   v_\mathrm{o}^\mathrm{n}$, with $0 \le n \le 3$.  The BLR gas exhibits highly super-solar chemical composition \citep{nagaoetal06,shinetal13}, so that assuming $Z = 5Z_{\odot}$\ is a reasonable choice for both [OIII] and CIV outflow, if the [OIII] emission is ascribed to a nuclear outflow.  Finally, to be consistent with the idea of a {\em bipolar} outflow, all the quantities in Table \ref{tab:php} have been multiplied by a factor 2.


It is interesting to make some considerations on the most likely values of the outflow parameters in very luminous quasars, and somehow constrain their upper limits.  The considerations below are focused on very luminous quasars such as the ones studied by the WISSH project and by \citet{sulenticetal17} and \citetalias{marzianietal16}. We will consider here  the 14 Pop. A objects of the ``HE'' sample of \citetalias{marzianietal16} and \citet{marzianietal16a} with $47.5 \lesssim \log L \lesssim 48.5$ erg s$^{-1}$. The relations of Table \ref{tab:php} are scaled to values typical of the HE sample.   In the redshift range $1 \lesssim z \lesssim 2.5$  where the population of most luminous quasars peaks, the angular distance is not increasing anymore with redshift \citep{hogg99}, implying a fairly constant scale around 8 Kpc/arcsec. For standard  ground based observations, with a slit width of  0.5 -- 1 arcsec,  all emission from  CIV is collected (obviously) and most or at least a significant fraction of [OIII] should be collected as well, although not necessarily all of it:   [OIII] emission may extend to the outer boundaries of optical galaxies and even beyond  (see, for instance, the impressive case of NGC 5252, \citealt{tadhuntertsvetanov89}).  


\subsection{CIV}  Estimating $L_{\rm line}$\ associated with unbound gas is not trivial, since the CIV emitting gas is probably still being accelerated by strong radiation forces within the BLR. As a lower limit, one can consider the fraction of the line that is already above a projected escape velocity. A more proper approach may be to consider that the gas is outflowing (i.e., the blue shifted component of Fig. \ref{fig:1}), and use a model in which gas cloud motions are accelerated under the effects of gravitation and radiation force (for example \citealt{netzermarziani10}). In this case the $v$\ entering the equations of the thrust and the kinetic power should be the terminal velocity $v_\mathrm{term}$, larger than the outflow velocity $v$ at  $r_{\rm CIV}$.  The CIV emitting region radius can be computed from the radius-luminosity relation derived for CIV, $r_{\rm CIV} \propto L^{b}$, with $b \approx 0.5 - 0.7$\   \citep[][see also contribution in the same research topic]{kaspietal07}.  While the density of the low-ionization part of the BLR is fairly well constrained \citep{negreteetal12,martinez-aldamaetal15}, the density of the outflowing component is not, although we can assume $8 \le \log n \le 10$\ [cm$^{-3}$], $0.2 \le L_\mathrm{line}/L_\mathrm{line,tot} \le 1$.  The thrust and the kinetic power may be larger than the values reported in Table \ref{tab:php} by a factor 10 and 100 respectively if radiative acceleration drives a wind with $k \approx 10$\ which may be the case for high Eddington ratio ($\gtrsim 0.7$, { following \citealt{netzermarziani10}}). 



\subsection{[OIII]} \label {oiii} For reasonable values of $r$, almost all of the blue shifted [OIII] emission should have escaped from the BH gravitational pull { ($k =1$)}. The full value of $L_{\rm line}$\ could be taken as a first guess of the outflowing gas luminosity. It is also reasonable to assume that the gas density is between the  [OIII] critical density   $\log n \sim 5.5$ [cm$^{-3}$] and the typical density of outer narrow line regions, $\log n \sim 3$ [cm$^{-3}$]. The distance $r$ remains a critical parameter   in the absence of  spatially resolved information. The ISAAC observations of the HE sample \citepalias{marzianietal16} were carried out with a slit width of 0.6 arcsec centered on the quasar: this implies that emission within $\approx$ 2.4 kpc was collected. Imposing mass conservation ($\dot M_\mathrm{[OIII]} \approx \dot{M}_\mathrm{CIV}$, and [OIII] emission at critical density  implies $r_\mathrm{[OIII]} \sim [5 \cdot 10^{9}/10^{5} ]^{\frac{1}{2}} \sim 10^{2.35} \sim 2 \cdot 10^{2}$ pc, if  $v_\mathrm{[OIII]}/v_\mathrm{CIV} \approx 5$, and $r_\mathrm{CIV} \approx 1$ pc.  An alternative assumption is motivated by  previous results on the low-$z$ blue outliers: it was possible to model both CIV and [OIII] with the same velocity field, assuming that the two lines were emitted with a velocity a constant factor { $ 1.5$ the local virial velocity i.e., slightly}  above the local escape velocity \citep[][cf. \citealt{komossaetal08}]{zamanovetal02}. Then $v_\mathrm{[OIII]}/v_\mathrm{CIV} = \sqrt{r_\mathrm{CIV}/r_\mathrm{[OIII}}$, if the factor remains the same for the two lines. We derive $r_\mathrm{[OIII]} \approx 25$\ pc.   This line of reasoning clearly emphasizes the necessity of obtaining spatially-resolved [OIII] data  with density diagnostics, and  of tracking the velocity field   as close as possible to the nucleus in prototypical cases that could help constrain observations lacking spatial resolution.


\subsection{Relation to luminosity and radiation thrust}

The average luminosity of the Population A HE sample sources is   $\approx 10^{47.8}$ erg/s. The average peak velocity shift of  the Pop. A CIV blueshifted component is $\approx -3000$ km/s. Typical $r_\mathrm{CIV}$ are around 1pc, and the typical CIV luminosity is $10^{45}$ erg/s. Even assuming $k = 10$,  $v_\mathrm{o} \approx 3000$ km/s,  the $\dot{\epsilon}$ value  is  several orders of magnitude below the bolometric luminosity: $\log \dot{\epsilon} \approx \log L - 2.4 $, a factor ten lower than the value of 5\%\ efficiency needed for a structural and dynamical effect on the host galaxy \citep[e.g., ][]{kingpounds15}. This limit might however be reached if the gas density is a factor $\approx 10$\ lower than assumed. Similar considerations apply to [OIII]: if the outflow is more compact than assumed in Table \ref{tab:php}, then an increase by a factor 10 -- 50 is possible.  However, with the values of Tab. \ref{tab:php} the estimates for $\log \dot{\epsilon}$ for [OIII] are four orders of magnitude  below $0.05 L$. 

The mechanical thrust values $\dot{M}v_\mathrm{o}$\ are also lower than   the radiation trust  $L/c \sim 10^{37.3}$ g cm s$^{-2}$. Again, $\dot{M}v_\mathrm{o}$\  may reach values of the same order or in excess by a factor of 20 of the radiation thrust \citep{zubovasking12} if $n$ is lower in the case of CIV and $r$ for [OIII] is $\ll 1$ Kpc.   A similar scenario involving $\dot{M}v_\mathrm{o}$\ and $\dot{\epsilon}$ too low at face values to explain the black hole --  bulge mass relation   was depicted by \citet{carnianietal15}.  Accepted at face values, these estimates suggest that, even in these very luminous quasars, the mechanical feedback estimated from mildly ionized gas may not be sufficient to reach the effect necessary for an evolutionary  feedback on the host galaxy, unless the outflow parameter are stretched to the limit of plausible values. Even if the required conditions are met, the observations  do not exclude  an important role of the active nucleus radiation force in driving the outflow.  In addition,  the estimates sample only one component of the nuclear outflow: the mildly ionized one which is, especially for the BLR, associated to a relatively small amount of matter. High ionization plasma, atomic and molecular gases are not considered, although even in the local Universe we have a spectacular example of massive molecular outflow, Mark 231 \citep[e.g., ][]{feruglioetal15}. 


\section{Conclusion}

Evidence of HIL blueshifts are ubiquitous, and at high luminosity they become impressive involving very large shifts in broad and narrow HILs. The mass outflow rates indicated by both [OIII] and CIV are extremely high, only somewhat lower than the accretion mass influx needed to sustain the observed luminosity for modest radiative efficiency ($\sim$100 M$_\odot$ yr$^{-1}$ at efficiency 0.1). Even in the most luminous quasars, it is not obvious whether powerful outflows can have the ability to disrupt the host galaxy gas. However, it is likely that [OIII] and CIV trace only a part of the mass outflow.   Accounting for  multiphase outflows will be one of the major challenges of present and future observational and theoretical studies.   



\section*{Funding} A.d.O.,  and M.L.M.A. acknowledge financial support from the Spanish Ministry for Economy and Competitiveness through grants AYA2013-42227-P and AYA2016-76682-C3-1-P. J. P. acknowledges financial support from the Spanish Ministry for Economy and Competitiveness through grants AYA2013-40609-P and AYA2016-76682-C3-3-P. D. D. and A. N. acknowledge support from grants PAPIIT108716, UNAM, and CONACyT221398.  



\bibliographystyle{frontiersinSCNS_ENG_HUMS} 







\end{document}